# Amorphous Nanoconfinement Enables Self-sustaining Sabatier Reaction at Ambient Conditions

Zhiyong Qiu[1], Cheng Li[2], Jinzhen Yang[1], Fangkun Sun[1], Zheng Zhang[3], Canwen Yu[1], Weizheng Cai[1], Liang Guo[3], Yutong Gong[4], Junjie Wang[4], Meng Danny Gu[2], Jiazhen Wu[1, 5]

[1] Department of Materials Science and Engineering, Southern University of Science and Technology, Shenzhen, 518055, China.

[2] Eastern Institute for Advanced Study, Eastern Institute of Technology, Ningbo, China.

[3] Department of Mechanical and Energy Engineering, Southern University of Science and Technology, Shenzhen 518055, China.

[4] State Key Laboratory of Solidification Processing, School of Materials Science and Engineering, Northwestern Polytechnical University, Xi'an, Shaanxi, 710072, China.

[5] State Key Laboratory of Quantum Functional Materials, Southern University of Science and Technology, Shenzhen, 518055, China.

**The Sabatier reaction, the catalytic hydrogenation of $CO_2$ into $CH_4$, offers a cornerstone for carbon capture and utilization, and *in-situ* resource utilization during space exploration; however, it faces a fundamental thermodynamic-kinetic paradox: although highly exothermic, conventional catalysts still require continuous external heating to activate $CO_2$ and maintain stable operation. Here we report an amorphous silica-embedded ruthenium catalyst that enables a long-term self-sustaining autothermal Sabatier reaction dispensing with external energy supply. Operating under ambient conditions, this system achieves a record-high $CH_4$ yield of 0.50 mol $g_{cat}^{-1}$ $h^{-1}$ with ~100% selectivity, stable operation for over 2,000 hours, and a record-low catalyst bed temperature down to ~100 °C. This exceptional self-sustaining behavior stems from the synergistic effect of the catalyst's ultralow effective thermal conductivity (0.27 W $m^{-1}$ $K^{-1}$), induced by amorphous nanoconfinement, and its superior intrinsic activity. This synergy generates localized hot spots at Ru sites while suppressing macroscopic heat loss. *In situ* measurements further reveal $CH_4$ formation even at 54 °C and identify a \*CO-mediated pathway for $CO_2$ methanation. The reaction ignites readily with a lighter or focused sunlight and persists even under forced convection from an electric fan, demonstrating strong environmental tolerance. By removing the need for constant energy input, this "ignite-and-forget" system paves the way for decentralized Power-to-Gas systems and autonomous fuel production in resource-constrained environments like Mars.**

## Main

The fossil fuel-dominated global energy system is facing severe challenges: massive consumption has accelerated resource depletion, while $CO_2$ emissions from combustion are the primary driver of global warming[1-3]. Since the Industrial Revolution, atmospheric $CO_2$ has risen by more than 140 ppm (from ~280 ppm to over 420 ppm)[4], intensifying extreme weather events and threatening sustainable development. In response, advancing renewable energy and $CO_2$ conversion technologies has become essential for achieving carbon neutrality and limiting warming to 1.5 °C[1-3,5,6]. At the forefront of these solutions is the Power-to-Gas (PtG) concept, which uses intermittent renewable sources such as wind and solar to generate green hydrogen that converts captured $CO_2$ into methane via the Sabatier reaction, enabling large-scale chemical storage of electricity[4,7-20]. The resulting methane offers a volumetric energy density ~3.2 times that of hydrogen and integrates seamlessly into existing natural gas infrastructure, alleviating global energy-distribution imbalances and buffering renewable intermittency[21]. By 2030, green-hydrogen-based $CO_2$ methanation is projected to yield tens of billions of cubic meters of synthetic methane annually. Moreover, the Sabatier reaction is indispensable for space exploration, powering water recycling on the International Space Station and enabling *in-situ* methane rocket fuel production on Mars[4,22]. Thus, $CO_2$ methanation research holds immense scientific significance for resolving today's energy–environment crisis and supporting future interplanetary societies.

The Sabatier reaction is strongly exothermic ($CO_2$ + 4$H_2$ → $CH_4$ + 2$H_2O$, $\Delta H$ = −165 kJ mol$^{-1}$). Thermodynamically, lower temperatures favor higher equilibrium $CO_2$ conversion and $CH_4$ selectivity[23]. However, the exceptional stability of the $CO_2$ molecule—characterized by its high C=O bond energy of approximately 800 kJ mol$^{-1}$—imposes a substantial kinetic barrier, rendering the reaction inefficient at low temperatures[24]. Consequently, conventional catalytic systems typically require elevated temperatures (> 300 °C) and pressures (> 1 atm) or additional external stimuli (such as light, plasma or electric field) to sustain $CO_2$ activation and achieve practical reaction rates[7,25-31]. This highlights a fundamental paradox: despite being highly exothermic and theoretically energy-self-sufficient, the process still relies heavily on continuous external energy supply. The resulting low energy efficiency, increased equipment complexity, and high operating costs pose significant challenges for integrating conventional methanation technologies with distributed renewable energy systems. Although recent progress in catalyst development and reactor design has reduced operating temperatures and minimized external energy demands[8,9,11,17,24,32-37], efficient and stable $CO_2$ methanation without any external energy input remains an elusive goal that demands further breakthroughs.

Here we show an amorphous silica (a-SiO$_x$)-confined Ru catalyst that enables fully self-sustaining $CO_2$ methanation using only $CO_2$ and $H_2$ as feedstocks, driven solely by the reaction enthalpy without any continuous external energy input (**Fig. 1a**). Under ambient conditions, the catalyst operates stably for over 2000 h, achieving 87% $CO_2$ conversion, near-unity $CH_4$ selectivity, and a methane space-time yield of 0.50 mol g$^{-1}$ h$^{-1}$—outperforming previously reported Ru-based systems that require external stimuli. The catalyst bed self-maintains at ~220 °C through reaction heat alone and remains stable even under forced cooling down to ~100 °C. *In situ* studies reveal the underlying mechanisms of this autothermal performance. Cold-start can be readily achieved with simple triggers such as a lighter or focused sunlight, and the system tolerates strong convective flow. This work overcomes the long-standing reliance of conventional methanation on continuous external energy input, offering a promising solution for distributed renewable-energy applications.

**Autothermal CO$_2$ methanation performance**

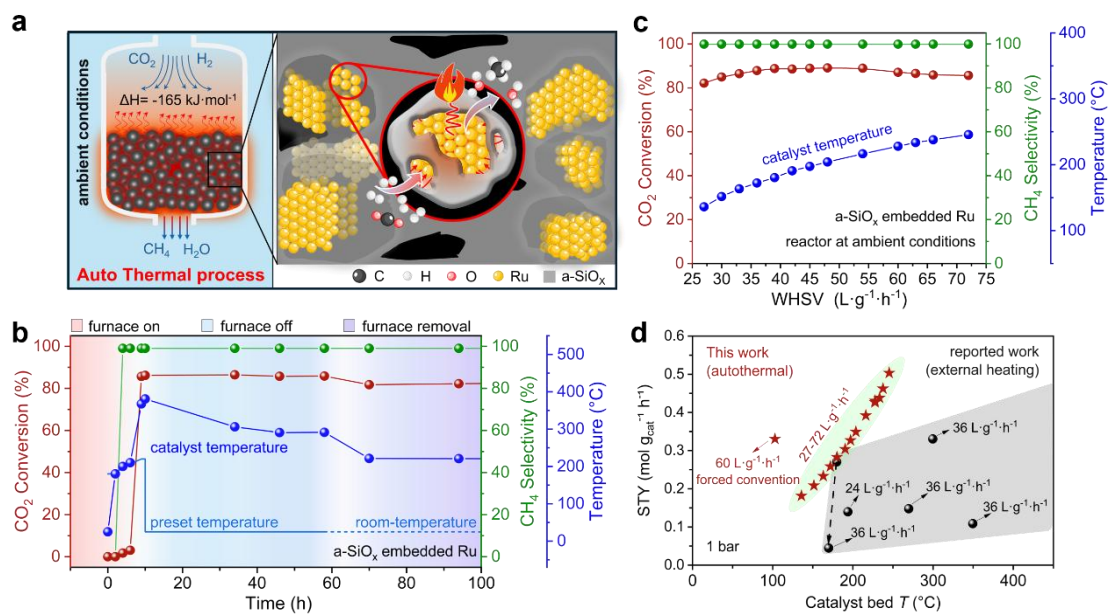

**Fig. 1 | Autothermal CO$_2$ methanation over an a-SiO$_x$ embedded Ru catalyst. a,** Schematic illustration of the autothermal CO$_2$ methanation process and the heat-retention structure enabled by a-SiO$_x$ nanoconfinement around Ru active sites. **b,** Time-on-stream profiles of CO$_2$ conversion, CH$_4$ selectivity, and catalyst bed temperature under different external thermal conditions. The catalyst maintains high activity and stable temperature independently of external heating. **c,** Effect of WHSV (reaction flow rate adjusted only) on equilibrium catalyst bed temperature and catalytic performance under ambient conditions. **d,** Methane STY for various catalysts. Standard reaction conditions: 100 mg catalyst, H$_2$:CO$_2$ molar ratio = 4:1; total flow rate = 100 mL·min$^{-1}$ (panel **b**) or varied as indicated (panels **c** and **d**).

The unique nanoconfinement structure of the a-SiO$_x$-embedded Ru catalyst provides two key advantages: markedly enhanced catalytic activity and effective locking of reaction heat through an amorphous thermal insulating layer around the active sites. This synergy enables fully self-sustaining CO$_2$ methanation under ambient temperature and pressure (**Fig. 1a**).

**Fig. 1b** shows the detailed autothermal performance of the Sabatier reaction. The catalyst was tested under different external thermal conditions (furnace on, furnace off, and furnace completely removed). When the furnace temperature approached the setpoint of 220 °C, the catalyst bed temperature rapidly increased to ~350 °C, demonstrating a strong reaction enthalpy-driven heating effect. After turning off the furnace (with the reactor still inside the furnace chamber), the bed temperature gradually decreased to ~290 °C. Upon complete removal of the furnace (i.e., under true ambient conditions), the temperature stabilized at ~220 °C (**Supplementary Video 1**). Notably, the catalyst maintained >80% $CO_2$ conversion and near-unity $CH_4$ selectivity even without any external heating. In contrast, control experiments with a 5 wt% Ru-$CeO_2$ and a 15 wt% Ru/SBA-15 catalyst under identical conditions showed that both catalysts failed to sustain the autothermal process once the furnace was turned off, with the reaction quickly quenching and the temperature dropping to room temperature (**Supplementary Figs. 1-4**).

The autothermal performance was further evaluated at different weight hourly space velocities (WHSV = 27–72 $L·g^{-1}·h^{-1}$) under ambient conditions (**Fig. 1c**). As the WHSV decreased, the equilibrium catalyst bed temperature gradually dropped from 246 °C (72 $L·g^{-1}·h^{-1}$) to 136 °C (27 $L·g^{-1}·h^{-1}$), confirming the reaction enthalpy-controlled behavior. Notably, $CH_4$ selectivity remained ~100% across the range, while $CO_2$ conversion stayed high (82.1–89.1%). The lowest temperature for the self-sustaining process was 102.5 °C, at which the catalyst still achieved 67.4% $CO_2$ conversion with ~100% $CH_4$ selectivity (**Supplementary Table 1**). The methane space-time yield (STY) was compared with literature values obtained under external heating (**Fig. 1d** and **Supplementary Table 2**). Remarkably, our catalyst achieved a high STY of 0.50 mol $g^{-1}$ $h^{-1}$ without any external energy input, outperforming all reported Ru-based thermal catalysts. Even at a bulk temperature approaching 100 °C, it maintained an impressive STY of 0.33 mol $g^{-1}$ $h^{-1}$.

## Catalyst characterization

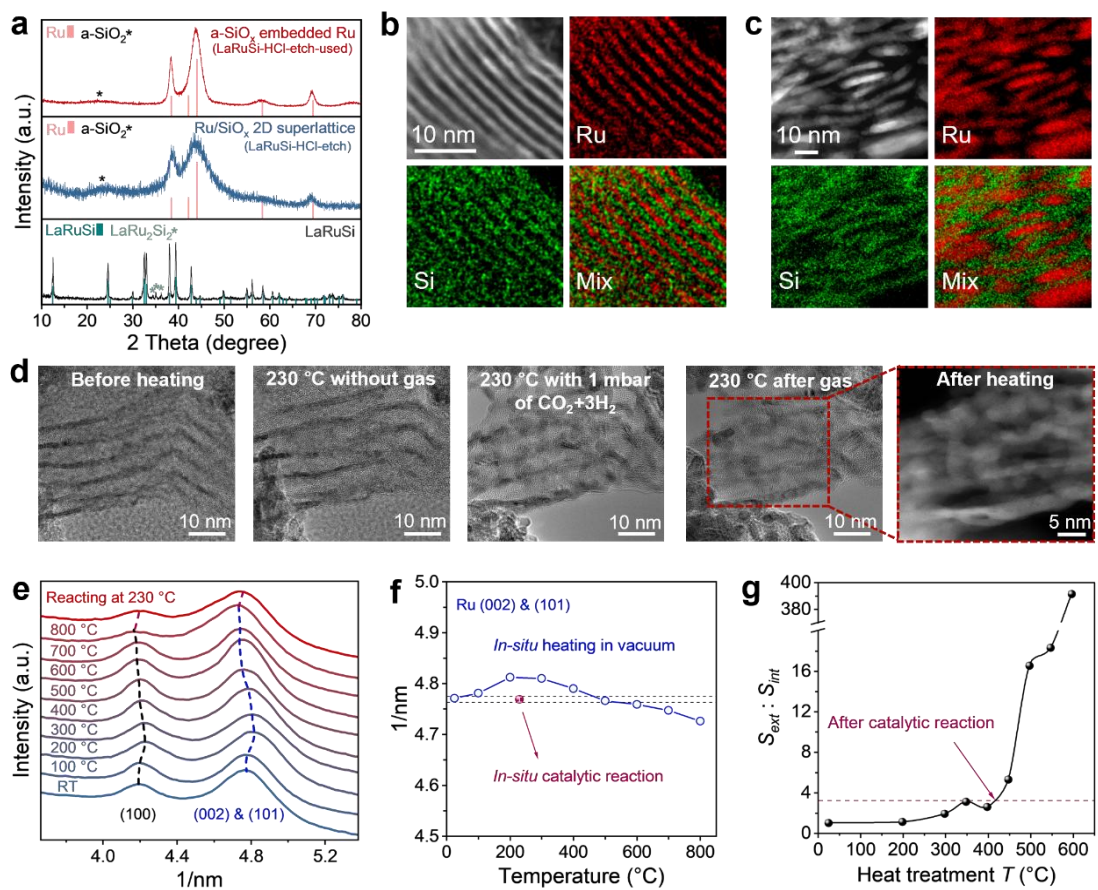

**Fig. 2 | Structural evolution and *in situ* characterization of the a-SiO$_x$-embedded Ru catalyst. a,** XRD patterns of the fresh and used catalysts. **b, c,** TEM images and corresponding EDS elemental maps of the fresh (**b**) and used (**c**) catalysts. **d,** Sequential *in situ* ETEM images captured at different stages of gas introduction, showing the dynamic structural evolution under operational conditions. **e,** Variation of the exposed Ru lattice spacing measured dynamically under *in situ* heating in the absence of reactant gas, compared with the post-reaction catalyst. **f,** Comparison of the interplanar spacings of Ru (002) under different conditions shown in (**e**). **g,** Ratio of external to internal surface area of the catalyst as a function of temperature, determined by the T-plot method. *In situ* ETEM experiments (**d**–**e**) were performed at 230 °C under a gas pressure of 1 mbar (H$_2$:CO$_2$ = 3:1).

The a-SiO$_x$-embedded Ru catalyst was prepared by selective HCl etching of the two-dimensional electride LaRuSi, followed by *in situ* structural reconstruction during catalysis. X-ray diffraction (XRD) showed that La was selectively removed upon etching, completely destroying the LaRuSi lattice and yielding broad diffraction peaks of metallic Ru and amorphous $SiO_2$ (**Fig. 2a**)[38]. After the catalytic reaction, the Ru Bragg peaks intensified, indicating further structural reconstruction and improved Ru crystallinity. Scanning electron microscopy (SEM), high-angle annular dark-field scanning transmission electron microscopy (HAADF-STEM), and energy-dispersive X-ray spectroscopy (EDS) confirmed the formation of a layered Ru/a-$SiO_x$ nano-sandwiched 2D superlattice structure after etching (**Fig. 2b and Supplementary Figs. 5-7**). Following catalysis, this nano-stripe structure reconstructed into Ru nanoparticles embedded within an a-$SiO_x$ matrix, resembling nuts set in a date cake (**Fig. 2c**). This morphology contrasts with that of the reference catalysts (Ru-$CeO_2$ and Ru/SBA-15; **Supplementary Figs. 2, 3**). High-resolution X-ray photoelectron spectroscopy (XPS) revealed that both Ru and Si were reduced after reaction, with Ru approaching a near-zero valent state and Si close to tetravalent (**Supplementary Fig. 8**).

*In situ* environmental transmission electron microscopy (ETEM) under reaction conditions directly visualized the dynamic structural evolution (**Fig. 2d, Supplementary Fig. 9, and Supplementary videos 2 and 3**). Heating the catalyst to 230 °C in vacuum produced no obvious morphological change. However, introduction of the reactant gas ($H_2$:$CO_2$ = 3:1) triggered rapid and drastic structural reconstruction, forming Ru nanoparticles spatially confined by a-$SiO_x$. This suggests that reaction enthalpy generates localized hot spots around the Ru active sites, driving the observed transformation.

To quantify the hot-spot effect, *in situ* control experiments were performed in vacuum using Ru lattice spacing as an internal thermometer (**Supplementary Fig. 10**). Systematic comparison of the Ru (100), (002) and (101) lattice spacings during catalysis at 230 °C with those measured in vacuum (**Fig. 2e, f**) showed that the lattice parameters closely matched those of a sample heated to 500 °C in vacuum, confirming substantial local overheating of the active sites.

Nitrogen adsorption–desorption measurements further corroborated the localized overheating effect (**Supplementary Table 3** and **Fig. 11**). The external-to-internal surface area ratio, determined by the T-plot method as a function of annealing temperature in Ar (**Fig. 2g**), increased markedly with temperature, reflecting progressive collapse of internal pores. The used catalyst after reaction exhibited a ratio nearly identical to that of a fresh sample annealed at 400–450 °C in Ar, confirming that the reaction generates localized hot spots in this temperature range. Notably, the catalyst retained a high specific surface area of 128.6 m² g⁻¹ even after annealing at 600 °C for 10 h, demonstrating excellent sintering resistance.

Integrating these multi-dimensional characterizations, we propose the following micro-mechanism. During the strongly exothermic $CO_2$ methanation, the rapid release of reaction heat at the Ru active sites creates transient high-temperature "hot spots". This localized thermal shock fractures the initial layered Ru structure, leading to its reorganization into nanoparticles, while simultaneously inducing localized migration and reconfiguration of the surrounding a-SiO$_x$ matrix. The resulting architecture functions as a nanoscale insulating blanket that locks in reaction heat for spontaneous autothermal operation while confining Ru species to prevent agglomeration and sintering.

**Heat dissipation analysis**

The amorphous silica nanoconfinement structure is expected to strongly influence the thermal conductance and dissipation of the catalyst. To quantify this effect, we analyzed the heat dissipation process of the reactor operating under true ambient conditions. A steady-state radial heat transfer model (**Fig. 3a**) was employed for the cylindrical catalyst bed (Methods). Three main pathways were considered: radial conduction within the catalyst bed, radial conduction through the quartz tube wall, and combined convective–radiative heat loss from the outer quartz wall to the ambient environment. The heat conduction contribution is described by

$$T(0) - T(r_2) = Q(R_{\text{cat}} + R_{\text{tube}}) = \frac{Q}{\pi L}\left[\frac{1}{4k_{\text{cat}}} + \frac{\ln(r_2/r_1)}{2k_{\text{tube}}}\right] \tag{1}$$

where $T(0)$ and $T(r_2)$ are the center and outer-wall temperatures, $Q$ is the reaction heat generation power, $R_{cat}$ and $R_{tube}$ are the thermal resistances of the catalyst bed and quartz tube, $L$ is the bed length, $k_{cat}$ and $k_{tube}$ are the effective thermal conductivities of the catalyst bed and fused quartz, and $r_1$, $r_2$ are the inner and outer radius of the tube. Under steady-state autothermal operation, $T(0) = 498.15$ K, $T(r_2) = 436.15$ K and $Q = 1.96$ W. Using the parameters in **Supplementary Table 4**, the calculated values are $R_{cat} = 30.07$ K W$^{-1}$, $R_{tube} = 1.95$ K W$^{-1}$ and $k_{cat} = 0.27$ W m$^{-1}$ K$^{-1}$. The catalyst bed dominates the total thermal resistance because of its much lower effective thermal conductivity compared with conventional supported catalysts[39].

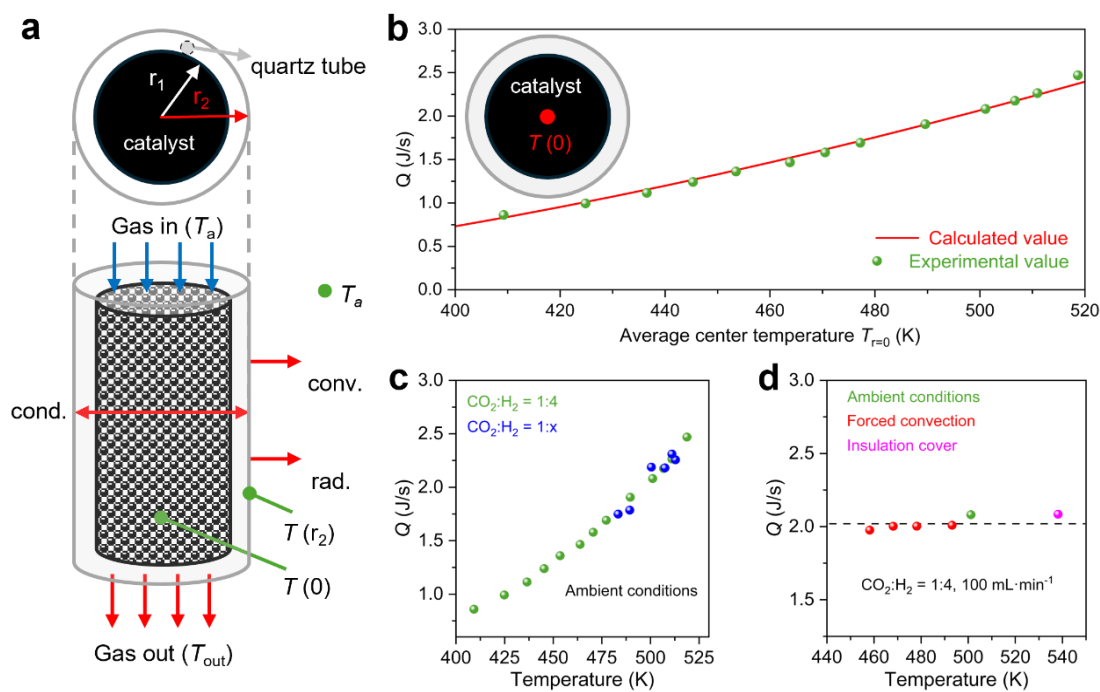

**Fig. 3 | Heat dissipation analysis of the reactor. a,** Schematic illustration of the heat transfer process, showing conduction, convection, and radiation boundaries. **b,** Relationship between catalytic reaction heat generation power ($Q$) and catalyst bed temperature ($T(0)$). **c,** Relationship between $Q$ and $T(0)$ under various feeding gas conditions. **d,** Relationship between $Q$ and $T(0)$ under various external cooling conditions.

The complete heat dissipation process, including convection and radiation, is

described by

$$T(0) - T_a = Q(R_{\text{cat}} + R_{\text{tube}} + R_{\text{c-r}}) = Q\left[R_{\text{cat}} + R_{\text{tube}} + \frac{1}{2\pi L(h+h_r)\left(r_2+\frac{r_1^2}{L}\right)}\right] \quad (2)$$

where $T_a$ is the ambient temperature, $R_{\text{c-r}}$ is the effective convective–radiative resistance, and $h$ and $h_r$ are the convective and radiative heat transfer coefficients. In the investigated temperature interval (~373 K), $R_{\text{cat}}$ and $R_{\text{tube}}$ are assumed to be unchanged, while $h$ and $h_r$ are temperature dependent (Methods). The relationships between $Q$ and $T(0)$ under different WHSV are shown in **Fig. 3b**. Equation (2) fits the experimental data with a goodness of fit of 0.995, and the fitted convective coefficient is consistent with the results obtained from the Churchill–Chu correlation, showing the validity of the theoretical model (Methods). Simulations based on the fitted parameters (**Supplementary Fig. 12**) show that the temperature decreases gradually from the catalyst centre to the outer shell, indicating that the average autothermal working temperature is lower than the measured centre value. The strong dependence of catalyst temperature on $k_{\text{cat}}$ highlights the critical role of low thermal conductivity in sustaining autothermal catalysis.

The autothermal process is remarkably tolerant to convection. Heat carried away by the outlet gas contributes only ~6.4% of the total dissipation (Methods). This robustness was further confirmed by deliberately varying the CO$_2$:H$_2$ feed ratio (**Fig. 3c**), which produced nearly identical trends in $Q$ and $T(0)$. Applying forced convection with an electric fan reduced the convective–radiative resistance $R_{\text{c-r}}$ from 33.02 K W$^{-1}$ to 16.93 K W$^{-1}$ and lowered the catalyst bed temperature from 493 K to 458 K; however, both the catalytic performance and the reaction heat generation power remained nearly constant (**Fig. 3d**). Adding a thermal insulating cover increased the bed temperature to 538 K without altering performance, further demonstrating the inherent stability of the autothermal regime.

## Catalytic mechanism

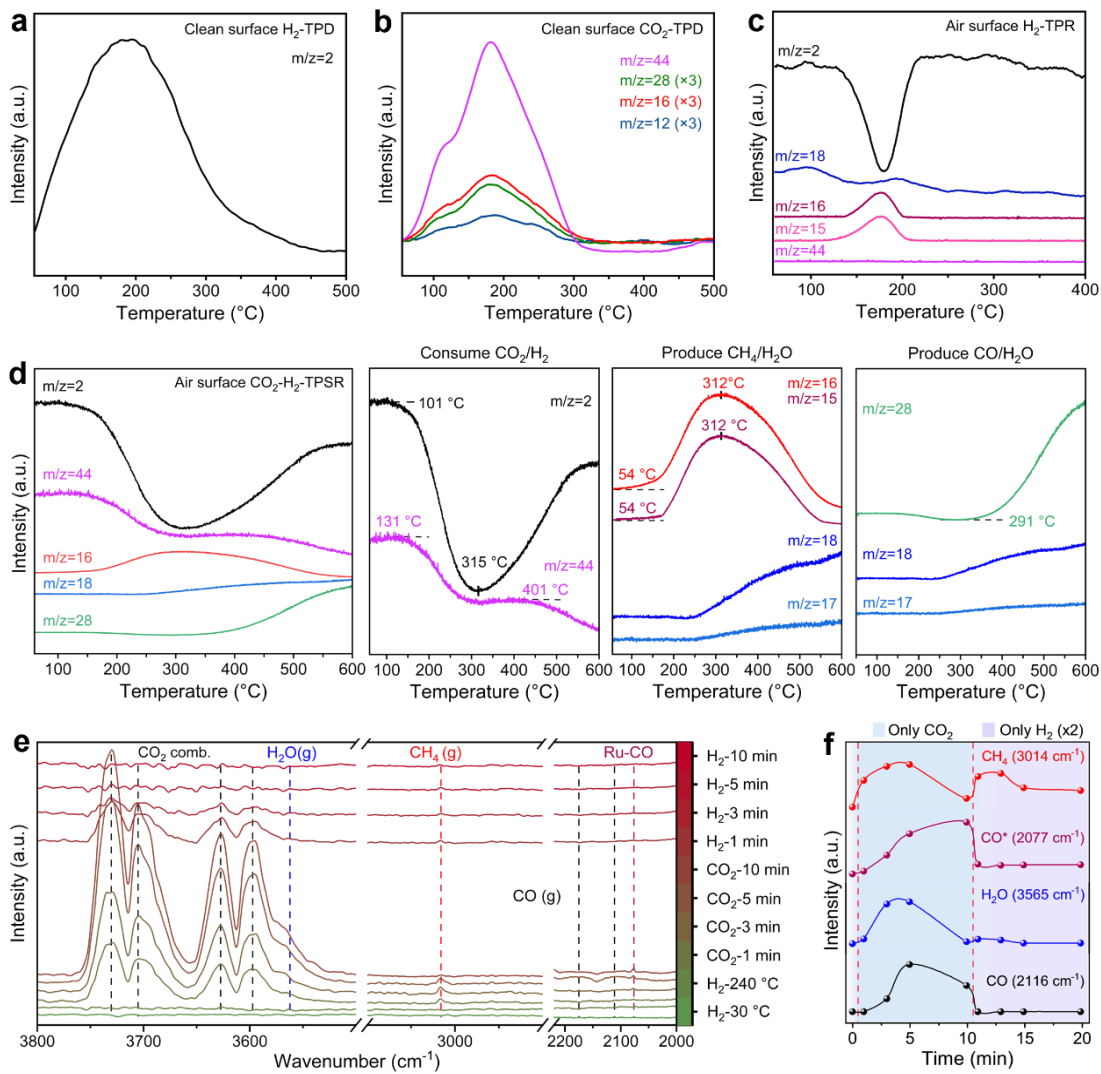

**Fig. 4 |** *In situ* **surface reaction analysis over the a-SiO$_x$-embedded Ru catalyst. a,** H$_2$-TPD profile of the clean surface. **b,** CO$_2$-TPD profile of the clean surface. **c,** H$_2$-TPR profile of the catalyst after air exposure. **d,** TPSR profile for CO$_2$ hydrogenation. **e,** Time-resolved *in situ* DRIFTS spectra recorded at 240 °C under varying feeding gas conditions. **f,** Evolution of the normalized intensities of key surface intermediates as a function of time, extracted from the spectra in **e**. The clean surface in **a** and **b** was obtained by sequential pretreatment of the catalyst at 400 °C under He flow for 2 h, H$_2$ flow for 2 h, and He flow for another 2 h to completely remove surface adsorbates.

Surface chemical properties and reaction intermediates were investigated by temperature-programmed desorption/reduction/surface reaction (TPD/TPR/TPSR) and *in situ* diffuse reflectance infrared Fourier transform spectroscopy (DRIFTS) (**Fig. 4**). The clean surface was obtained by sequential pretreatment at 400 °C under He, $H_2$ and He flows for 2 h each (**Supplementary Fig. 13**). $H_2$- and $CO_2$-TPD measurements show that both gases are adsorbed at 50 °C and begin desorption below 100 °C (**Fig. 4a, b**), demonstrating the catalyst's superior low-temperature adsorption and activation capacity. Notably, the catalyst can adsorb $CO_2$ directly from ambient air (**Supplementary Fig. 13b, c**). $H_2$-TPR experiments performed on an air-exposed catalyst (bearing surface $CO_2$) exhibit clear methanation signals (**Fig. 4c**), whereas no such signals appear on a clean surface (**Supplementary Fig. 14**), indicating the potential for ambient-temperature $CO_2$ activation.

TPSR measurements reveal that $CH_4$ formation begins at an exceptionally low temperature of ~54 °C, with the signal intensity increasing continuously and peaking near 312 °C (**Fig. 4d**). The generation of $CH_4$ is accompanied by the consumption of $CO_2$ and $H_2$, with similar peaking temperatures (~315 °C), directly indicating the $CO_2$ methanation reaction. The onset of $CO_2$ and $H_2$ consumption occurs at higher temperatures (101 °C and 131 °C, respectively) than the $CH_4$ formation onset, likely due to the low signal-to-noise ratio at low conversions, where minor consumption signals are masked by large background signals. Above 291 °C, the reverse water-gas shift pathway becomes prominent, producing increasing amounts of CO while the $CH_4$ signal declines.

*In situ* DRIFTS further elucidated the reaction pathway (**Fig. 4e, f** and **Supplementary Fig. 15**). After $H_2$ pretreatment at 400 °C, the catalyst was cooled to room temperature and heated to 240 °C under $H_2$. Upon switching to $CO_2$, characteristic $CH_4$ peaks (3014 and 1304 cm$^{-1}$)[40,41] and gas-phase water (1300–2000 cm$^{-1}$)[40] appeared within 1 min, indicating that pre-stored active hydrogen (H*) rapidly hydrogenates $CO_2$. Even after cutting off gas-phase $H_2$, these stored H* species sustain $CO_2$ methanation for more than 5 min, consistent with the strong $H_2$ adsorption observed in TPD (**Fig. 4a**). A distinct peak at 2077 cm$^{-1}$, assigned to Ru(CO*)$_n$ intermediates[42], accumulates with prolonged $CO_2$ exposure; gaseous CO (2105 and 2177 cm$^{-1}$) [43] appears after ~3 min when the surface H* pool is depleted. Upon switching back to $H_2$, the Ru-CO* signal vanishes within 1 min while $CH_4$ peaks intensify rapidly. No other stable intermediates were detected throughout the transient process. These observations establish that adsorbed CO is the key reaction intermediate: $CO_2$ dissociates to form linear CO*, which is progressively hydrogenated by H* (from pre-stored or gas-phase $H_2$) to yield $CH_4$.

**Catalytic stability**

To assess the practical applicability of the catalyst, we evaluated its autothermal performance under long-term continuous operation, multiple activation–shutdown cycles, varying feed gas ratios, and different ignition strategies (**Fig. 5**).

The catalyst exhibits outstanding long-term stability (**Fig. 5a**). During over 2000 h of continuous operation under ambient conditions, both $CO_2$ conversion and $CH_4$ selectivity remained constant, delivering methane at a steady rate of ~1 L h$^{-1}$. Post-reaction weighing confirmed the absence of carbon deposition. Reusability tests (**Fig. 5b**) further demonstrate that the catalyst retains its structural integrity and catalytic performance across repeated activation and shutdown cycles.

For distributed Power-to-Gas applications involving fluctuating renewable hydrogen supply, adaptability to variable feed compositions is essential. As shown in **Fig. 5c**, the catalyst maintains high activity and near-unity $CH_4$ selectivity across a wide range of $H_2/CO_2$ molar ratios. Under hydrogen-rich conditions ($H_2/CO_2$ = 6.33), it achieves complete $CO_2$ conversion with 100% $CH_4$ selectivity. The resulting $H_2/CH_4$ gas mixture can be directly used as a fuel[44-46], making it suitable for many practical applications without further purification.

The self-sustaining nature of the system also enables highly flexible, low-energy ignition. The autothermal process can be rapidly initiated using simple localized heat sources, including a conventional lighter (**Fig. 5d**), an industrial heat gun (**Fig. 5e**), or focused sunlight (**Fig. 5f**). These straightforward activation methods highlight the system's suitability for resource-limited and decentralized settings.

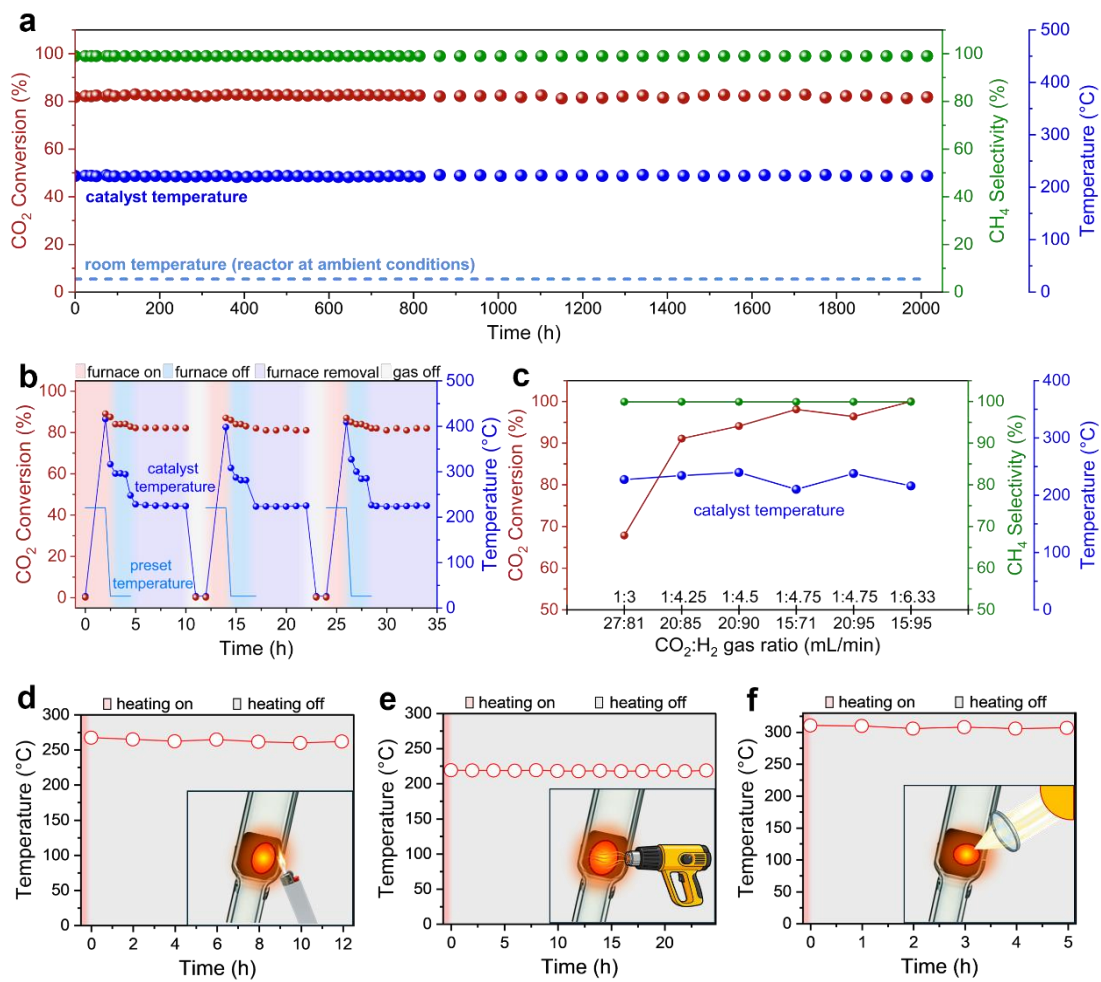

**Fig. 5 | Adaptability of the autothermal catalytic system under various conditions. a,** Long-term stability test under ambient conditions for over 2000 h. **b,** Cyclic performance assessment. **c,** Dependence of catalytic performance ($CO_2$ conversion and $CH_4$ selectivity) and catalyst bed temperature on the $CO_2/H_2$ feed ratio. **d–f,** Demonstration of versatile ignition strategies using a lighter (**d**), an industrial heat gun (**e**), and concentrated sunlight (**f**). The observed differences in catalyst bed temperature are due to the different experimental environments: carbon black coverage from incomplete combustion of the lighter in (**d**), operation in a fume hood in (**e**), and exposure to sunlight in (**f**). All autothermal experiments were performed with 100 mg catalyst and a total feed gas flow rate of 100 mL·min$^{-1}$ ($H_2:CO_2$ = 4:1), unless otherwise specified.

**Conclusion**

In conclusion, we have developed an amorphous silica-embedded Ru catalyst that achieves fully self-sustaining $CO_2$ methanation driven solely by reaction enthalpy. The system demonstrates exceptional long-term stability exceeding 2000 hours under ambient conditions, delivering more than 80% $CO_2$ conversion, near-unity $CH_4$ selectivity, and a high methane space-time yield of 0.50 mol $g^{-1}$ $h^{-1}$, while self-maintaining bulk temperatures down to ~100 °C even under strong convective cooling. Its remarkably simple activation and excellent tolerance to fluctuations in inlet gas flow rates and ratios further enable operation under real-world variable conditions.

This outstanding autothermal behavior originates from the amorphous nanoconfinement effect, which functions like a thermal "cotton jacket" around the Ru active sites, dramatically reducing effective heat dissipation while enhancing intrinsic catalytic activity and preventing metal sintering. These synergistic advantages provide a powerful, generalizable strategy for designing energy-self-sufficient catalysts in exothermic systems. By resolving the long-standing thermodynamic–kinetic paradox of the Sabatier reaction, this work paves the way for decentralized Power-to-Gas technology on Earth and autonomous *in-situ* fuel production on Mars, transforming an elusive goal into a practical reality for carbon-neutral energy storage and interplanetary exploration.

## Methods

### Chemicals

All chemicals were used as received without further purification. Hydrochloric acid (36–38 wt%) was purchased from Shanghai Lingfeng Chemical Co., Ltd. Lanthanum (99.9%) and silicon (99.999%) were obtained from Thermo Scientific, and ruthenium (99.95%) was sourced from PrMat Co., Ltd.

### Materials preparation

LaRuSi was synthesized by arc-melting stoichiometric amounts of lanthanum, ruthenium, and silicon ingots under an argon atmosphere. The melting process was repeated five times to ensure homogeneity, with weight loss below 0.1%. The resulting silver-colored ingots were ground into powder using an agate mortar in an Ar-filled glovebox. Since the as-melted LaRuSi was not single-phase, the powder was annealed at 1000 °C for 10 days to eliminate impurity phases. The LaRuSi powder (~0.25 g) was chemically etched to selectively remove lanthanum by immersion in 100 mL of 3 M HCl solution in a centrifuge tube. The mixture was stirred for 12 h, after which the solid was separated by centrifugation, washed five times with distilled water, and vacuum-dried at room temperature for 12 h. The a-SiO$_x$-embedded Ru catalyst was formed *in situ* from the HCl-etched LaRuSi during the catalytic reactions.

### Materials and catalytic characterizations

Powder XRD patterns were collected on a Bruker D2 Phaser diffractometer using Cu K$\alpha$ radiation ($\lambda$ = 1.5418 Å). Sample morphology was examined by field-emission SEM (TESCAN MIRA3). The Brunauer–Emmett–Teller (BET) surface area of the catalysts was determined from nitrogen adsorption–desorption isotherm measurements at –196 °C using a Micromeritics Tristar 3000 gas adsorption instrument. Prior to the measurements, the catalysts were heat-treated at 200 °C. Surface chemical information was analyzed by XPS on an ESCALAB Xi+ photoelectron spectrometer (Thermo Scientific) with a monochromatic Al K$\alpha$ X-ray beam (1486.6 eV). All binding energies were referenced to the C 1s peak at 284.8 eV.

HAADF-STEM images of the samples were captured on a double Cs-corrected FEI Themis G2 thermal-field emission microscope equipped with a probe Cs-corrector operating at 300 kV. A convergence angle of 25 mrad and a collection angle range of 38 to 200 mrad were used for imaging. *In situ* ETEM characterization was performed using a Titan ETEM G2 microscope operated under a controlled gas atmosphere of 1 mbar $CO_2/H_2$ (1:3 ratio). During heating experiments, dynamic processes were recorded at a frame rate of 2 frames per second. For static imaging, samples were heated at a rate of 20 °C min$^{-1}$ to the target temperature, followed by a 10-minute equilibration period to eliminate thermal drift prior to image acquisition.

$CO_2$-TPD, $H_2$-TPD, $H_2$-TPR, and $CO_2$ hydrogenation TPSR measurements were all carried out on a BELCAT-A instrument (BEL Japan, Inc.). The instrument was equipped with a thermal conductivity detector (TCD) and an online mass spectrometer (Bell Mass, BEL Japan, Inc.) for qualitative and quantitative analysis of gas-phase species in all tests. For the clean surface TPD measurements, the sample was pretreated by sequential He flow for 2 h, $H_2$ flow for 2 h, and He flow for another 2 h at 400 °C to completely remove surface adsorbates. Prior to TPD measurements, the sample was further treated in $CO_2/H_2$ at 50 °C for 2 h. For TPSR measurements, the gas conditions were 2 mL·min$^{-1}$ $CO_2$ and 80 mL·min$^{-1}$ 5% $H_2$/Ar.

*In situ* DRIFTS experiments were performed on a Bruker INVENIO S FT-IR spectrometer (spectral range 400–4000 cm$^{-1}$, resolution 4 cm$^{-1}$, 64 background scans, 32 sample scans). The catalyst was pretreated via in situ reduction at 400 °C for 1 h under 28 mL·min$^{-1}$ $H_2$, followed by Ar purging (30 mL·min$^{-1}$) at 400 °C for 2 h. The sample was then cooled to room temperature under Ar, purged for an additional 30 min, and a background spectrum was recorded. Transient tests were conducted sequentially as follows: the sample was held at 30 °C under 28 mL·min$^{-1}$ $H_2$ for 10 min, then heated to 240 °C under the same $H_2$ flow. Spectra were collected at 1, 3, 5, and 10 min after switching to 7 mL·min$^{-1}$ $CO_2$, followed by spectra collection at 1, 3, 5, and 10 min after switching back to 28 mL·min$^{-1}$ pure $H_2$ at 240 °C.

**Catalytic reaction**

Catalytic performance for $CO_2$ methanation was evaluated in a fixed-bed quartz reactor at ambient pressure. The reaction temperature was monitored by a thermocouple with its tip placed directly in contact with the catalyst bed. Typically, 100 mg of catalyst was loaded into the reactor, *in situ* reduced at 400 °C for 2 h under a 40 mL·min$^{-1}$ $H_2$ flow, and then cooled to 25 °C. A reactant mixture containing 20% $CO_2$ in $H_2$ was introduced at a total flow rate of 100 mL·min$^{-1}$, corresponding to a weight hourly space velocity (WHSV) of 60 L·g$^{-1}$·h$^{-1}$. The effluent gas was passed through an ice-water trap to remove water and analyzed online using an Agilent 8890 gas chromatograph equipped with a thermal conductivity detector (TCD), a 5 Å molecular sieve column, and two Hayesep Q columns. All data were collected after the system had stabilized for at least 1 h at each temperature to ensure steady-state conditions.

$CO_2$ conversion ($X_{CO_2}$) and $CH_4$ selectivity ($S_{CH_4}$) were calculated according to the following equations:

$$X_{CO_2} = \frac{n(CH_4)_{out} + n(CO)_{out}}{n(CH_4)_{out} + n(CO_2)_{out} + n(CO)_{out}} \qquad (3)$$

$$S_{CH_4} = \frac{n(CH_4)_{out}}{n(CH_4)_{out} + n(CO)_{out}} \qquad (4)$$

where $n(CH_4)_{out}$, $n(CO)_{out}$ and $n(CO_2)_{out}$ represent the molar flow rates of $CH_4$, CO, and $CO_2$ in the reactor outlet stream, respectively.

The space-time yield of $CH_4$ was calculated as:

$$\text{STY}(CH_4) = \frac{RF \times V\% \times X_{CO_2}}{V_m \times m_{cat}} \quad (\text{mol} \cdot g^{-1} \cdot h^{-1}) \qquad (5)$$

where RF is the total flow rate of the reaction feed gas, $V\%$ is the volume fraction of $CO_2$ in the inlet gas, $V_m$ is the molar volume of an ideal gas at 25 °C and 1 atm (24.5 L mol$^{-1}$), and $m_{cat}$ is the mass of the loaded catalyst.

**Heat dissipation analysis**

1. Construction of the steady-state radial heat transfer model

To simplify the analytical solution while retaining physical validity, four fundamental assumptions were made based on the actual reaction conditions and reactor geometry. (1) Steady-state operation: Once the reaction reaches a stable self-sustained state, the temperature at any point in the catalyst bed is time-invariant and the total energy of the system is balanced. (2) Radial heat transfer: The reactor is an axisymmetric cylindrical structure; thus, temperature gradients along the circumferential and axial directions are negligible, and only radial heat conduction is considered. (3) Uniform internal heat source: The catalyst bed is uniformly packed, so the volumetric heat generation rate from the methanation reaction is assumed to be evenly distributed throughout the bed. (4) Negligible gas enthalpy change: The enthalpy change of the inlet and outlet gas stream accounts for only a very small fraction of the total heat released by the methanation reaction and can therefore be ignored in the energy balance.

2. Estimation of the enthalpy change of the inlet and outlet gas

Inlet conditions: pressure $P = 1$ bar, inlet temperature $T_{in} = 298$ K, total gas flow rate $V_{total} = 100$ mL·min$^{-1}$, H$_2$:CO$_2$ molar ratio = 4:1.

Outlet conditions: pressure $P = 1$ bar, outlet temperature $T_{out} = 380$ K, CO$_2$ conversion = 82% (long-term stability test).

Based on the stoichiometry of the Sabatier reaction and 82% CO$_2$ conversion, the molar flow rates of the outlet components were calculated as:

$$n_{CO_2} = 1.471 \times 10^{-4} \text{ mol} \cdot \text{min}^{-1}, \quad n_{H_2} = 5.885 \times 10^{-4} \text{ mol} \cdot \text{min}^{-1},$$

$$n_{H_2O} = 1.341 \times 10^{-3} \text{ mol} \cdot \text{min}^{-1}, \quad n_{CH_4} = 6.703 \times 10^{-4} \text{ mol} \cdot \text{min}^{-1}.$$

The temperature difference of the gas stream is $\Delta T_{gas} = 82$ K. The molar

enthalpy change of each component is given by

$$\Delta H_{m,i} = C_{P,m,i} \times \Delta T_{gas} \tag{6}$$

where $C_{P,m,i}$ is the average constant-pressure molar heat capacity of component $i$ over 298–500 K (values taken from the NIST Chemistry WebBook[47] and listed in **Supplementary Table 5**.

The total enthalpy change of the gas stream is

$$\Delta H_{gas} = \sum n_{i,out} \cdot \Delta H_{m,i} \tag{7}$$

Substituting the data yields $\Delta H_{gas} = 0.125$ W. The molar reaction enthalpy of the Sabatier reaction at 225 °C (498.15 K) was corrected to $\Delta_r H_m(495.15 \text{ K}) = -175.50$ kJ mol$^{-1}$ using Kirchhoff's law, corresponding to a total reaction heat release rate $Q_{reaction} = 1.96$ W. Therefore, the ratio of gas enthalpy change to total reaction heat release is $\Delta H_{gas}/Q_{reaction} \approx 6.4\%$. This confirms that the gas enthalpy change is negligible compared with the total reaction heat release and has no significant impact on the core results of the steady-state radial heat transfer model.

3. Estimation of the effective radiative and convective heat transfer coefficients

The total heat dissipation from the reactor consists of radiative and natural convective heat transfer from the outer wall of the quartz tube to the ambient environment. Both pathways are incorporated into the steady-state model via series thermal resistance superposition.

The effective radiative heat transfer coefficient is derived by treating the outer quartz wall as a diffuse-gray surface in an infinite ambient environment (view factor ≈ 1) and applying the Stefan–Boltzmann law. Using the mean-temperature approximation (valid because the maximum temperature difference between $T(r_2)$ (400–520 K) and $T_a$ (298 K) is much smaller than their arithmetic mean), the simplified expression is

$$h_r(T) \approx 4\varepsilon\sigma \left[\frac{T(r_2) + T_a}{2}\right]^3 \tag{8}$$

where $\varepsilon = 0.74^{48}$ (average emissivity of the quartz tube) and $\sigma = 5.67 \times 10^{-8}$ W m$^{-2}$ K$^{-4}$ (Stefan–Boltzmann constant).

The effective convective heat transfer coefficient is expressed as $h(T) = A \times T(r_2) - B$, where $A$ and $B$ are fitting coefficients. Substituting this expression together with equation (8) into equation (2) and fitting the steady-state data in **Fig. 3b** yields

$$h(T) = 0.178 \times T(r_2) - 43.96 \tag{9}$$

The validity of the fitted $h(T)$ was verified using the curvature-corrected Churchill–Chu correlation for natural convection[49] around a short vertical cylinder ($L = 9.8$ mm):

$$Nu_L = \left\{0.825 + \frac{0.387 Ra_L^{1/6}}{[1+(0.492/\text{Pr})^{9/16}]^{8/27}}\right\}^2 \times \left\{1 + 0.300\left[32^{0.5} \cdot Gr_L^{-0.25}\left(\frac{L}{D}\right)\right]^{0.909}\right\} \tag{10}$$

where $Nu_L$ is the Nusselt number based on cylinder height $L$, $Ra_L$ is the Rayleigh number, and Pr is the Prandtl number ($\approx 0.7$, evaluated at the average temperature of the wall and ambient air)[50]. The Rayleigh numbe[50] is calculated as $Ra_L = Gr_L \cdot \text{Pr}$, where the Grashof number is

$$Gr_L = \frac{g\beta(T(r_2) - T_a)L^3}{v^2} \tag{11}$$

(with $g = 9.8$ m·s$^{-2}$, $\beta = 1/T$ (average temperature), and $v = 1.93 \times 10^{-5}$ m$^2$·s$^{-1}$)[51]. At $T(r_2) = 436$ K and $T_a = 298$ K, the calculated values are $Gr_L = 9.33 \times 10^3$, $Ra_L = 6.53 \times 10^3$, $Nu_L = 6.36$, and

$$h = \frac{Nu_L \cdot k_{\text{air}}}{L} \tag{12}$$

(with $k_{\text{air}} \approx 0.034$ W m$^{-1}$ K$^{-1}$)[50]. The theoretical convective coefficient is ~22.1 W m$^{-2}$ K$^{-1}$, which is in the same order of magnitude as the fitted value of 33.65 W·m$^{-2}$·K$^{-1}$. The slightly higher fitted value is reasonable and can be attributed to minor air disturbances in the laboratory environment (e.g., fume hood ventilation). This agreement confirms the reliability of the fitting result.